\documentclass[MSNbibl,nameyear,dvips]{arxstspdf}
\usepackage{flushend}
\usepackage{stfloats}

\volume{26}
\issue{2}
\pubyear{2011}
\firstpage{238}
\lastpage{239}
\doi{10.1214/11-STS331REJ}
\referstodoi{10.1214/10-STS331}

\makeatletter
\def\bsuffix #1{#1}
\makeatother

\begin{document}
\begin{frontmatter}

\title{Rejoinder}
\runtitle{Rejoinder}
\pdftitle{Rejoinder}
% of Discussion on Bayesian Models and Methods in
%Public Policy and Government Settings by S. E. Fienberg

\begin{aug}
\author{\fnms{Stephen E.} \snm{Fienberg}\corref{}\ead[label=e1]{fienberg@stat.cmu.edu}
\ead[label=u1,url]{http://www.stat.cmu.edu/fienberg/}}

\runauthor{S. E. Fienberg}

\affiliation{Carnegie Mellon University}

\address{Stephen E. Fienberg is Maurice Falk University Professor,
Department of Statistics, Heinz College, Machine Learning Department, and Cylab, Carnegie Mellon University, Pittsburgh, Pennsylvania 15213-3890, USA
\printead{e1,u1}.}

\end{aug}

% ABSTRACT

% KEYWORDS

\end{frontmatter}

The three discussants have offered three complementary perspectives on
the material in my paper and in different ways help to sharpen the
focus on the appropriateness and utility of the Bayesian perspective in
government and policy settings.   I am indebted to them for their
comments and critiques, which by and large remain couched in
compliments, for which I also thank them!

I did consider responding using a variation on Alan Zaslavsky's clever
culinary metaphor.  But it would be difficult to match him tit for tat
as he was even able to  adapt  Jimmie Savage's (\citeyear{Sav61})
oft-repeated remark that the Fisherian  fiducial school's approach was
``a bold attempt to make the Bayesian omelet without breaking the
Bayesian eggs,''  to apply to some modern frequentists who borrow from
Bayesian ideas.  In the end, I decided to simply offer a few
observations  of why I think so much has changed over the past 50
years, with the hope that these might explain why I differ with a~number of the comments from the discussants.

My education as a statistician goes back to the early 1960s when the
number of people expressing strong Bayesian perspectives could fit in a
small seminar room at a university, and we often did so as part of the
Seminar in Bayesian Econometrics that the late Arnold Zellner convened
twice a year.  Applications in those days typically meant small-scale
numerical illustrations using conjugate priors for analytical
convenience, and Bayesian approaches were rarely taught in statistical
courses except for at a~handful of places, and then only to graduate
students.  The towering achievement of Mosteller and Wallace
(\citeyear{MosWal64})  in bringing a systematic Bayesian\vadjust{\eject} approach to
the analysis of the Federalist Papers thus served as an eye-opener to
the statistical community and showed that Bayesians could do serious
substantive applications that harnessed the power of the largest
computers of the time.  For some insights into their effort I recommend Chapter 4 of Mosteller's~\citeyear{Mos10}
posthumously-published autobiography on this work.\vadjust{\goodbreak}

For most of today's readers of \textit{Statistical Science}, it may be
hard to imagine the almost complete dominance of the frequentist
perspective in our journals and in application fifty years ago.  It was
in part for this reason that I began my examples  with some details on
the NBC Election Night Forecasting team from the 1960s because it too
was an anomaly.   On the other hand, something that was true in the
1960s, as it is today, was that most statistical education and research
was built around statistical models and inference from them.  The
principal departure from this model-based perspective  came in the area
of sample surveys, where essentially the only source of random variation
considered by authors and practitioners was that associated with the
random selection of the sample and this then provided the basis for
inference about population quantities---what we now describe as
design-based inference.    This perspective was so deeply embedded in
the operations of national statistical agencies that it still remains
through to today.  I remember making a  presentation in the late 1970s
at a sample survey symposium on why one should view surveys on crime
victimization in the context of longitudinal models for individual
respondents and households, in which I criticized the narrow
cross-sectional perspective adopted by the U.S. Census Bureau in its
work on the National Crime Survey (which was in fact a longitudinal
survey but not analyzed as such).  My remarks were barely completed
when Morris Hansen, who was seated in the front row, stood and took me
to task because I did not understand the limitation of my perspective
and the fact that government agencies understood the limitations of the
data they collected and why models had no place in their analysis.%\looseness=-1

Even in the 1950s and 1960s, frequentists were being influenced by
Bayesian ideas, and Charles Stein's results on shrinkage estimation,
which were later\vadjust{\eject} adapted in the form of empirical Bayesian estimation
by Efron and Morris (\citeyear{EfrMor73}), drew heavily on the form of Bayesian
weighting of sample quantities with prior ones, albeit with a
frequentist outcome in mind.  Several of us taught this Bayesian
motivation to students at the University of Chicago, where I was a
faculty member from 1968 to 1972, and I suspect this may have
indirectly influenced Bob Fay, who was my undergraduate advisee and who
later co-authored with\vadjust{\goodbreak} Roger Herriot  their landmark paper on
small area estimation (Fay and Herriot, \citeyear{FayHer79}).

The foregoing is a somewhat longwinded way of explaining why, to
paraphrase the Virginia Slims commercial from the 1960s, ``we've come a
long way,  baby,''  Graham Kalton's protestations to the contrary.
When I first visited the Census Bureau, shortly after my exchange with
Morris Hansen,  few of the statisticians could even understand my ideas
on log-linear models and their relevance to census activities.  This
has changed quite markedly, although many in the agencies have strong
training in statistical theory and methodology, the remains resistance to explaining the model-based and often Bayesian
motivation of approaches being advocated, even\break though there is internal
recognition of this strong influence.  This is especially true in the
context of post-enumeration surveys for assessing censal accuracy.
Where Kalton and I disagree strongly is on the use of models to analyze
and interpret the results of large-scale government social surveys. For
example, most of the interesting analyses of the Current Population
Survey (CPS), used in the U.S. to produce the monthly unemployment
rate, are based on statistical models and on the implicit longitudinal
structure of the survey.  This is certainly the perspective of most
policy analysts outside the government who use the CPS in their work.

Zaslavsky notes that one of the aspects of the objective Bayesian
school is its use of Bayes as a~device to generate calibrated
(frequentist) probability statements.  That is clearly a substantial
part of the modern  literature, but it should play little role in many
applications, I believe.  Consider disclosure limitation to protect
confidentiality in statistical databases.  We are surely interested
less in protecting an infinite sequence of hypothetic databases
generated using the same probabilistic mechanism than we are in
protecting the database at hand, once we have collected it.  Thus
conditioning on the data we have rather than the data we might have had
makes eminently more sense to me.  If an objective prior at the top of
a hierarchical model can succeed in doing this, I certainly have no
objections.

 Zaslavsky also refers to the practice of Bayesian model averaging.  Again
 this is a place where we do not fully agree.  I see Bayesian model averaging as
 fitting well within the subjective Bayesian paradigm, but primarily for
 prediction-like\vadjust{\goodbreak} problems where different models could conceivably have quite
 different and possibly non-overlapping specifications.  When model averaging
 is used for inference about parameters in models, however, the results are often
 nonsensical, because the ``same parameter'' in different models often has a~totally
 different meaning depending on the rest of the model specification.  Regression
 analysis offers a good example of this phenomenon.

David Hand notes that all models are approximations at best, and both
he and Graham Kalton refer to George Box's famous \textit{dictum} that
``all models are wrong, but some are useful.''   I agree and I also
agree with Hand that any approach to the analysis of data in practice
requires much more than invoking the Bayesian mantra.  Statisticians
really need to know what they are doing, both substantively and
statistically.  A Bayesian algorithm does not necessarily make for a
good Bayesian analysis and proper inferences.  Hand's example of
borrowing strength in large sparse tables of adverse reactions in the
post-marketing surveillance of drugs harks back to many of the earlier
examples of Bayesian ideas and methods I refer to in the paper.  I
thank him for this and the other examples.

All three discussants caution that Bayesian methods are not the answer
to all policy problems.  I~agree, and I~often adopt likelihood-based
methods in my own work when a \textit{de novo} Bayesian approach seems
forbidding.  I remain a subjective Bayesian, however, and no longer see
the ``threat'' of subjective priors as a major obstacle to the adoption
of Bayesian methods and analyses.

Does one size fit all?  Of course not.  But Bayesians come in all
stripes and varieties today and they work on diverse applications.
I believe we can look forward to the increasing use of Bayesian methods
in many domains, including those described in my paper.

% imsref loaded by svajune.rapalyte, 2011-05-09 09:31:33

\end{document}